\documentclass{article}

\usepackage{orcidlink}
\usepackage{PRIMEarxiv}
\usepackage{academicons}
\usepackage[utf8]{inputenc} 
\usepackage[T1]{fontenc}    
\usepackage{hyperref}       
\usepackage{url}            
\usepackage{booktabs}       
\usepackage{amsfonts}       
\usepackage{nicefrac}       
\usepackage{microtype}      
\usepackage{lipsum}
\usepackage{fancyhdr}       
\usepackage{graphicx}       
\graphicspath{{media/}}     

\title{Secured Fiscal Credit Model: Multi-agent Systems and Decentralised Autonomous Organisations for Tax Credit’s Tracking}

\author{
  \href{https://arxiv.org/search/cs?searchtype=author\&query=De+Gasperis,+G}{Giovanni De Gasperis}\orcidlink{0000-0001-9521-4711}, \href{https://arxiv.org/search/cs?searchtype=author\&query=Facchini,+S+D}{Sante Dino Facchini}\orcidlink{0000-0002-2009-5209} \\
  Department of Information Engineering, Computer Science and Mathematics\\
  University of L'Aquila, Italy\\
  \texttt{giovanni.degasperis@univaq.it, santedino.facchini@graduate.univaq.it} \\
   \And
  \href{https://arxiv.org/search/cs?searchtype=author\&query=Letteri,+I}{Ivan Letteri}\orcidlink{0000-0002-3843-386X}\\
  Computer Science Department \\
  Sapienza University of Rome, Italy\\
  \texttt{ivan.letteri@uniroma1.it} \\
}

\begin{document}
\maketitle

\begin{abstract}
Tax incentives and fiscal bonuses have had a significant impact on the Italian economy over the past decade. In particular, the "Superbonus 110" tax relief in 2020, offering a generous 110\% deduction for expenses related to energy efficiency improvements and seismic risk reduction in buildings, has played a pivotal role.

However, the surge in construction activities has also brought about an unfortunate increase in fraudulent activities. To address this challenge, our research introduces a practical system for monitoring and managing the entire process of the Superbonus 110 tax credit, from its initiation to redemption. This system leverages artificial intelligence and blockchain technology to streamline tax credit management and incorporates controllers based on a Decentralised Autonomous Organisation architecture, bolstered by a Multi-agent System.

The outcome of our work is a system capable of establishing a tokenomics framework that caters to the needs and functionalities of both investors and operators. Moreover, it features a robust control system to prevent inadvertent errors like double spending, overspending, and deceitful practices such as false claims of completed work.

The collaborative approach between the Decentralised Autonomous Organisation and the Multi-agent System enhances trust and security levels among participants in a competitive environment where potential fraudsters might attempt to exploit the system. It also enables comprehensive tracking and monitoring of the entire Superbonus process.

In the realm of engineering, our project represents an innovative fusion of blockchain technology and Multi-agent Systems, advancing the application of artificial intelligence. This integration guarantees the validation, recording, and execution of transactions with a remarkable level of trust and transparency.

\end{abstract}


\section{Introduction}
\label{sect:intro}

Italian energy policies and laws have a deep root in European Union guidelines, these mainly aim to improve the performance and to reduce the wastefulness of households' energy production systems. During the last years, the debate around energy efficiency rose in importance and is now central to determining the Next Generation EU funding policies\footnote{{\tiny\url{https://energy.ec.europa.eu/topics/energy-strategy\_en}}, last access Feb 2023}. As an EU Member, Italy has taken steps to encourage the energy efficiency increase in residential buildings. A program of tax deductions and fiscal bonuses, originally introduced in 1997 \cite{RePEc:bdi:wptemi:td_423_01} was relaunched in 2011 and renewed over the years till 2019. Finally, during 2020, the actual system of 110 percent Superbonus tax credit has been introduced (Art.119 DL Rilancio 34/2020)\footnote{{\tiny \url{https://www.gazzettaufficiale.it/eli/id/2020/05/19/20G00052/sg}}, last access Feb 2023}. The lawmaker instituted an unprecedented program that enables residential owners to complete energy efficiency upgrades and structural improvements to existing buildings without any upfront costs but rather through a tax credit mechanism.

The Superbonus at 110\% represents an opportunity for citizens and a valuable boost to the country’s economic revival. It is also turning into an obstacle race for thousands of users and condominiums, who have decided to take advantage of the incentive to renovate buildings and homes. Furthermore, it gives rise to speculation and in some cases outright fraud (\cite{euractivURL}).

Our contribution is to prevent Fraud \& Misuse by tracing the credit history and identifying all participants involved in the process. In fact, identity fraud and credit allocation to fraudulent recipients are the primary risks associated with the Superbonus 110 process. Another risk is the incorrect allocation of credit, for instance, a client may request a higher amount than they are owed or more properties than they are allowed. In all of these situations, tracing the assignment of every euro of the tax credit is of critical importance.

In this work, we present \textit{Secured Fiscal Credits Model} (SCFM), a framework that uses two Distributed Autonomous Organisations (DAOs) to track and control the whole Superbonus 110 tax credit process. The first one (Investors DAO) is meant to group the investors, collect all the values need to back up the tax credit system and tokenize such values. The second one (Operator DAO) will model the market of operators generating and spending credits in the Superbonus 110 administrative context.

Our proposal builds up from the integration of Artificial Intelligence (AI) and blockchain technology for tax credit management with a decentralised platform, like the SingularityNET Project \footnote{\url{https://singularitynet.io} Last access July 2023}, 
which allows the implementation of a decentralised marketplace for AI services integrated through a blockchain-based network \cite{goertzel2023artificial}.

Our aim is to achieve stability in the value of each token introduced in the Superbonus circuit to induce a 1:1 correspondence to the underlying fiat currency without actually exchanging any amount, allowing only to use of tokens among members of the DAO. Furthermore, we aim to control and track each token transaction in order to certify and backtrack all spending on services and goods involved between parties in the included smart contracts. 

Through the integration of the Multi-agent system with two Distributed Autonomous Organisations, we managed to implement the token minting and transferring of tax credits among the operators. The DAO\&MAS combination enables highly automated, unbiased management of value exchange among users through decentralised governance, without requiring full trust among users. As in (\cite{DyoubCLL21}),(\cite{DyoubCLG20}) was a MAS demonstrated to be able to monitor the ethical behaviour of dialogue systems in different domains.

The rest of this work is organised as follows: in Section \ref{sect:rw} are described the background and the related works. In Section \ref{sect:methodology}, we report the methodology applied followed by the scenarios considered and the system design in Section \ref{sect:scenarios} and in Section \ref{sect:sysarc} respectively. Finally, the conclusion and future works are given in Section \ref{sect:concl}.

\section{Background and Related Works}
\label{sect:rw}

\subsection{The Blockchain}
\label{sect:blockchain}
Tax Audit is an examination of records to determine whether a taxpayer has correctly reported its tax liabilities. \cite{Karakostas2021Kiayias} discuss the problem of facilitating tax auditing assuming ``programmable money'', i.e. digital monetary instruments that are managed by an underlying Distributed Ledger Technology (DLT). 

A distributed ledger is a type of database that keeps multiple copies of information in different nodes in the network, which can be updated consistently. This allows a full copy of the shared ledger, verifiable by the interleaved chain of data blocks and cryptographically signed maintaining integrity and availability by a protocol of consensus (\cite{nakamoto2008bitcoin},\cite{Gervais2016Karame}).

Nowadays, blockchain is considered an institutional innovation for the economic coordination of organisations; through a scalable blockchain, in fact, is possible to increase the productivity of existing processes by lowering transaction costs and avoiding costly intermediations \cite{Catalini2020Gans}. Such crypto-economic systems can provide an institutional infrastructure that facilitates a wide range of socio-economic interactions to influence participants in their behaviour \cite{Voshmgir2019Zargham}. 


Most notable, the Ethereum blockchain \cite{Buterin}, made it possible for Turing-complete code pieces named \textit{smart contracts} to be executed on a blockchain as tasks of a virtual processor machine. Smart contracts allow formalising interaction rules between blockchain transactions and digital workflows to coordinate economic activity and ensure that automated processes run according to predefined rules and state changes in the DAOs. 

A DAO is considered a multi-criteria decision-making problem that deals with evaluating a set of possible alternatives (see \cite{Baninemeh2021Farshidi}), so to present a decision model to participants that can possibly converge to a deliberation consensus \cite{Farshidi2018Jansen}.

In this paper, we focus on blockchain as a possibility to scale scenarios on the Superbonus 110\% tax credit's tracking previously exposed in \cite{DeGasperis2022Facchini}.   
    
\subsection{The Superbonus 110\%} 
\label{sb110}
In this section, we describe the Superbonus 110\% program and the procedure for claiming the discount on the invoice and the accrued credit from taxable income. They can be obtained - as per current legislation - by submitting specific documentation to demonstrate compliance with the requirements and proper execution of works. This includes an energy performance certificate, a statement that the work carried out meets certain technical standards and an asseveration of accounting and bookkeeping of the whole process. Furthermore, the execution of the works must meet specific requirements and steps. The whole project has to improve the energy footprint of the house and the response of the building structure to earthquakes.

\subsubsection{The Actual Legislation} 
Superbonus 110\%, unlike other tax allowances, provides for a higher-than-investment rate of deduction, as well as a different way of allowances, and claims.

The actual legislation disposes that for every 100 euros spent on renovation works, the House Owner will be entitled to a 110 euros tax deduction to be used over the next 5 years and divided into 5 equal annual instalments. Since the annual deduction is discounted from the taxpayer's gross tax (IRPEF), it is therefore recoverable within the limit of such an amount and cannot be carried forward or claimed back. Any deductible allowance exceeding the gross tax would be lost. A thorough and updated source of info on Superbonus 110 regulation can be found on the \textit{Agenzia delle Entrate} Italian website \footnote{{\tiny \url{https://www.agenziaentrate.gov.it/portale/web/guest/superbonus-110\%25}}, last access Feb 2023}. 

\subsubsection{Discount on the Invoice and The Accrued Credit} 
The legislator provided an alternative solution to the direct deduction with a discount on the invoice received for works. It is possible to transfer/sell the accrued credit to a third party, so, the Customer (House Owner) is automatically entitled to transfer the deductions gained by interventions allowed by Superbonus 110\%, also when has not enough tax to pay. It is possible to opt for a discount on the invoice of the total amount (i.e. 100\%) applied directly by the Supplier/General Contractor and corresponding to the maximum amount to be paid for the works VAT included. Then, the Customer will transfer in turn the tax credit (i.e. 110\%) to the invoice issuer.

At this point, the Supplier/General Contractor can set off the discounted invoice applied to the purchaser as a tax credit or transfer it to other financial intermediaries or credit institutions. In particular, he will also have at his disposal an extra 10\% calculated on the discounted invoice amount. This amount has been estimated by the Government to cover financial costs related to the overall project.

The \textit{Agenzia delle Entrate}, in order to apply the Superbonus 110\%, refers to the contract signed between the Customer (the taxpayer commissioning the work) and the construction company (General Contractor). The prices to be used for such works are specifically determined in the official price lists fixed quarterly in each region (\textit{Prezziari}).

\subsubsection{The Procedure}\label{sect:procedure} 
\begin{figure}[ht]
\centerline{\includegraphics[width=1.00\hsize]{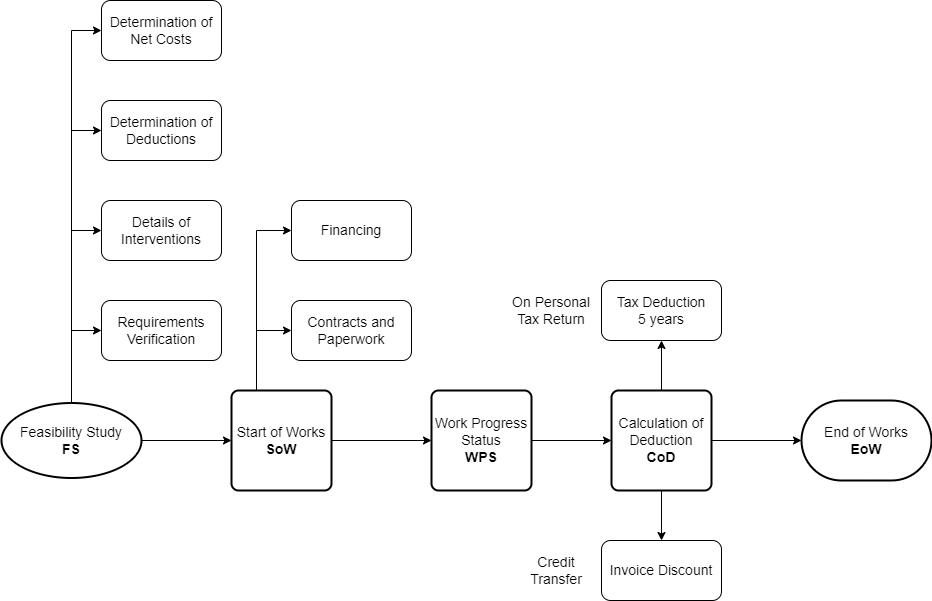}}
\caption{Time sequenced phases for the acquisition of Superbonus 110. (i) Starting with the FS by assessing costs, deductions, interventions, and verification. (ii) SoW definition of funding and contracts of the project. (iii) WPS paperwork and reports on the progress status of building interventions. (iv) CoD tracking and calculation of payments to be deducted as tax credits. (v) EoW closing of all works and asseverations, with a record of the intervention and archiving of documents.}
\label{fig:steps}
\end{figure}

There are four key steps for carrying out interventions under the Superbonus 110\%: First, the Feasibility Study to determine if the property qualifies. Second, the Start of the Work after permits are obtained. Third, the Calculation of Deductions to identify cost savings. Finally, the Closing of the Work completes the process. Figure \ref{fig:steps} shows the main steps along with their sub-options. An example of the credit request workflow which involves the relations with the Actors is exposed in figure \ref{fig:fase1}, while a detailed distribution and timing of works steps are shown in figure \ref{fig:sal}.

\begin{itemize}
    \item \textit{Step 1 - Feasibility Study (FS)} consists of three steps: (i) Verification that the envisaged intervention qualifies for relief such as building permits and urban planning compliance. (ii) Detailing of the hypothesised interventions and calculation of the deductions due (e.g., estimated metric calculation, detailed estimates). (iii) Determination of net investment cost (i.e., actual expenditure credit).

    \begin{figure}[ht]
    \centerline{\includegraphics[width=0.5\hsize]{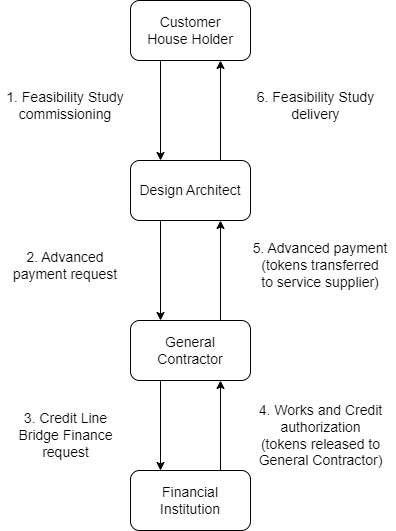}}
    \caption{Example of interaction between Actors involved in the credit request process. Boxes represent the actors, while the arrows have numbered actions that connect ordered relations and/or documents to be created between them.}
    \label{fig:fase1}
    \end{figure}

    \item \textit{Step 2 - Start of the Work (SoW)} represents the final balance of what was budgeted in the feasibility study. There are two different scenarios: (i) The client who has financial means and can therefore afford to wait until the end of the works to be able to accrue the tax credit. (ii) The client who needs to finance the works, with the accrual of the tax credit occurring in steps during the execution of the works (\textit{Work Progress Status}). This second option is the one happening most of the time, the General Contractor usually manages to set up the bridge finance with banks/financial institutions.

    \item \textit{Step 3 - Work Progress Status (WPS)} is the accounting act functional to the payment of the work done till that moment, it summarises all the works and all the supplies carried out from the beginning of the contract up to the day of issue. A copy of the lists of prices (\textit{prezziari}) is attached to the WPS and is checked constantly against the real expenditures to avoid overspending. The Director of Works is responsible for keeping each voice of cost within the limits defined by law (\textit{massimali}).

    \item \textit{Step 4 - Calculation of Deduction (CoD)} The annual deduction is recoverable up to the limit of the personal gross tax (IRPEF). There are two possible scenarios: (i) if the client has sufficient gross tax to absorb the annual deduction, it will be recovered in the tax return; (ii) if the tax is not sufficient, the client may opt for the discount on the invoice.
    For example, a contribution in the form of a discount on the invoice, up to a maximum amount equal to the consideration, is to be requested by the supplier. Then, the supplier recovers it by accruing a credit with the Tax Agency. This credit will be used for offsetting or in turn assigned to third parties (i.e. the deduction due is assigned to third parties (banks, insurance companies, post office, intermediary, other companies, individuals).

    \item \textit{Step 5 - End of the Work (EoW)} All the works are checked and tested, and a final balance of the intervention is drafted. Both Design Architects and Tax Auditors certify expenses and congruity on technical and fiscal sides. Once all these activities are performed the documents are deposited to the Italian Tax Agency which may check them within 7 years.
    \begin{figure}[ht]
    \centerline{\includegraphics[width=1.00\hsize]{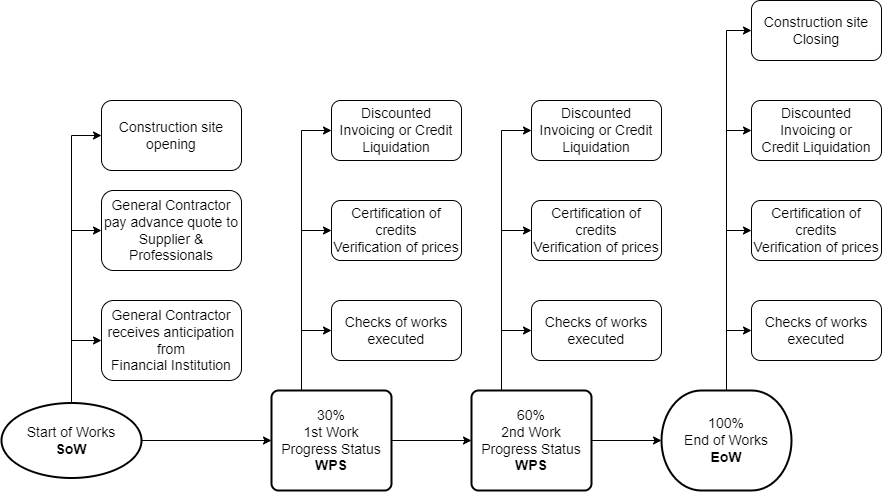}}
    \caption{Typical Work Progress Status distribution. The lower boxes represent the phases of work. The upper boxes are the actions/sub-phases requested for each main phase of the Workflow.}
    \label{fig:sal}
    \end{figure}
    
    As shown in figure \ref{fig:sal}, there are 3 WPSs, (i) the first is stated at 30\% of work, (ii) the second at 60\%, and (iii) the remaining one at the completion of the entire work (100\%). 
    
    By calculating the average time to carry out the work, it is possible to make a projection of the state of the WPS. Thus, the money required may be reserved over the foreseen time period (on average 8 months per building site for \textit{ecobonus+sisma}-bonus \footnote{Sisma-bonus is regarding the building enhancement towards seismic vulnerabilities} or 6 months for \textit{single eco}-bonus). This will correspond, as better specified in Section \ref{sec:42} and Section \ref{sec:53}, to freezing the correspondent amount of tokens in the Investors DAO.
    This behaviour is based on the principle of a constant progression of the accounts (Italian Chartered Accountants principle OIC 12 \footnote{{\tiny \url{https://www.fondazioneoic.eu/wp-content/uploads/2011/02/2022-05-OIC-12-Composizione-e-schemi-del-bilancio1.pdf}}, last access Feb 2023} see \cite{quagli2017bilancio} for a comprehensive analysis), the activities of checking the facts producing expenditure must be carried out at the same time as they occur and, therefore, must proceed in parallel with execution.
    
\end{itemize}

\subsection{Decentralised Autonomous Organisation (DAO)}
DAOs are a type of organisation that is controlled entirely by computational algorithms known as smart contracts which determine the rules of how the parties involved in the DAO cooperate. 
The idea of DAOs was outlined for the first time during the late 90s and was connected to the application of MAS to intelligent home sensors \cite{Dilger}. The first transposition of a real company on the blockchain was defined with the introduction of the Digital Autonomous Corporations concept (DACs) (see \cite{iceis15}). Later the introduction of the bitcoin time-chain (\cite{nakamoto2008bitcoin}) - which defines a public ledger with transparent transactions - and in 2015 the Ethereum blockchain allowed the tokenisation of assets and fully automated and incorruptible smart contracts. So a real governance process of an organisation became possible (\cite{Buterin}). 
DAOs take on the characteristics of the underlying DLT layers, including decentralised control, security through asymmetric cryptography keys, and the ability for smart contracts to self-execute.
DAOs are not bound to any particular regulation or law due to their decentralised nature. For these reasons, DAOs are natural candidates to represent real-world companies and complex organisations in blockchain environments (\cite{Sun2022Stasinakis}). DAO internal architecture comprises 4 major mechanisms, described as follows: (i) \textit{Smart Contracts}, (ii) \textit{Consensus Protocol}, (iii) \textit{Tokens} and (iv) \textit{Blockchain.}
The combination of these four elements is what makes it possible for a DAO to properly run at all times.

\section{Methodology}
\label{sect:methodology}

In this section, we define our methodology, which combines the knowledge base of constraints, agent design, and scenarios to develop the decision support system. Our method is split into two steps: (i) definition of a constraints knowledge base to capture domain knowledge, including objectives, and preferences of decision makers. This knowledge base also includes data on available resources, technologies, and regulations. (ii) Designing agents to represent decision-makers, stakeholders, and other entities involved in the problem. Agents are programmed to communicate with each other and negotiate based on the knowledge base and scenarios. 

\subsection{Constraints Knowledge Base}
\label{sub:CKB}
The \textit{Decreto Aiuti n. 50/2022} \footnote{\url{https://www.gazzettaufficiale.it/eli/id/2022/05/17/22G00059/sg}, last access Feb 2023} introduced the rules for calculating the percentage of intervention achieved. When assessing the WPS of 30\%, therefore, the overall intervention must be taken into account, including works carried out on the dwelling that do not fall within the scope of the Superbonus 110, for example, design costs, certifications and other professional fees. 
The term ``\textit{total intervention}'' is intended to define all expenses arising from the general economic framework of the intervention being carried out, including technical expenses relating to the services performed in connection with the works carried out on the building.

In the following are listed the six constraints designed, then we explored the characteristics of design constraints, types of constraints, and examples of constraints involving safety, cost, resources, and environmental impact. 

\newpage

\begin{itemize}
    \item \textit{Constraint 1} There is a specific agent (Director of Work) that verifies compliance with the original schedule plan. Starting from a monthly forecast, the agent will be able to estimate the maturity of the WPS based on the previously mentioned criterion about the distribution of WPS. This translates into the fact that possible states of the agents (Workflow Agent see \ref{sect:SFCMAS}) are the following: $$ x_i=\{Open, Anticipation, Sal1, Sal2, Eow, Archived\},$$ or in brief $ x_i=\{Op, Ant, S1, S2, Eow, Arc\} $. 
    We can at this point formalise the following expressions for each value of the state variable:
    
    $$x_i(0)\Longleftarrow Op\land \lnot Ant \land \lnot S1 \land \lnot S2 \land \lnot Eow \land \lnot Arc$$

    $$x_i(1)\Longleftarrow \lnot Op\land Ant \land \lnot S1 \land \lnot S2 \land \lnot Eow \land \lnot Arc $$ 
    
    $$\\ ... \\$$
    
    $$x_i(5)\Longleftarrow \lnot Op\land \lnot Ant \land \lnot S1 \land \lnot S2 \land \lnot Eow \land Arc$$

The KB constraint can be at this point defined as follows:

\begin{center}
$ true\Longleftarrow x_i(0)\lor x_i(1) \lor x_i(2) \lor x_i(3) \lor x_i(4) \lor x_i(5) $ 
\end{center}

    \item \textit{Constraint 2} On the basis of previous WPSs it is necessary to check financial expenditures and future needs at the closing of the WPS and thus verify that the expected demand for tokens in the Operators DAO ($dao_o$) does not exceed that forecast in the Investors DAO ($dao_i$). So if $d_t$ is the demand for the token and $f_t$ is the forecast need for a token at time $t$, we must have:
    $$d_t \leq f_{t-1}$$
       
    \item \textit{Constraint 3} Do not spend more than allocated. Electronic invoices will be the source of the flow of economic transactions, they allow a transparent view of all expenditures in the WPS, so all payments done ($p_s$) by the Workflow agent must not exceed the payments received ($p_r$)as anticipation for that state. Both must be made by traceable means. If $w_t$ is the state of the Workflow agent at time t:
    $$p_s(w_t) \leq p_r(w_{t-1})$$

    \item \textit{Constraint 4} The Homeowner may have at most two concessions per building unit to be refurbished. This translates in the condition that each Client ($c_i$) may not have more than 2 Workflow agents ($w$) opened:
    $$|w(c_i)|\leq 2$$

    \item \textit{Constraint 5} Each project (i) must be certified by both technical and financial asseveration ($a$) raised by chartered engineers ($eng$) and accountants ($acc$) to grant the fairness of the expenses, with respect to the interventions listed in the WPS. Thus if engineer (j) and accountant (k) are hired to asseverate project (i):
    $$true\Longleftarrow a(eng_{j})_{i}\land a(acc_{k})_{i}$$

    \item \textit{Constraint 6} The Contractor (GC) may have more than one construction site assigned. The maximum limit is determined either by the tax credit that may be handled or the value set in the SOA certification he has in place (e.g., 1 million euros works for private buildings renovation category). So if the GC has $j$ works of value $w_j$ each:
    $$\sum w(GC_i)_j \leq SOA(GC_i)$$
\end{itemize}

\subsection{Agents Design: Actors, Roles, and Actions}
\label{sect:32}
The Agents are the actors with their own roles and execute actions. The first step is to identify the environment and the entities involved in the process. We spotted two main areas: (i) the \textit{investors group} which includes all actors getting financial burdens and benefits in the process, and (ii) the \textit{operators group} that in turn aggregates all actors involved. Table \ref{tab:actors-actions} shows, in the DAO column, how some actors may take both groups' properties.

\begin{table}[!ht]
\begin{tabular}{|p{0.15\textwidth}|p{0.27\textwidth}|p{0.51\textwidth}|}
\hline
\multicolumn{1}{|c|}{\textbf{DAO}} & \multicolumn{1}{c|}{\textbf{Actor}} & \multicolumn{1}{c|}{\textbf{Actions}} \\ \hline
Investors  & Investor  & Invest money to buy credits and benefits from its selling   \\ \hline
Investors Operators   & Financial Inst. & Banks that buy/sell credits and lend money to Operators   \\ \hline
Investors Operators   & Customer   & House owner Sells/Transfers tax credits from works   \\ \hline
Operators  & General Contractor   & Manages work gets credits from Customers, sells to Financial Institutions    \\ \hline
Operators  & Sub-contractor  & Hired and paid by General Contractor  \\ \hline
Operators  & Supplier & Hired and paid by General Contractor      \\ \hline
Operators  & Design Architect  & Hired and paid by General Contractor  \\ \hline
Operators  & Tax Auditor & Hired and paid by General Contractor        \\ \hline
\end{tabular}
\caption{Relationship between the Actors and their affiliation DAOs in comparison to the user actions.}
\label{tab:actors-actions}
\end{table}


In the \textbf{Operators group}, we have all the actors with operational functions on the construction site as well as in the design and consultancy area. They use the Operators DAO to receive and make payments for goods and services and to exchange and certify documents.
In the \textbf{Investors group} we have all the actors with a passive interest in the Superbonus process, in particular professional and non-professional investors as well as in the financial entities. They use the Investors DAO to invest their money and make a profit. The actors on the Investors DAO are the following:

\begin{itemize}
    \item \textbf{Investors:} In this category are included all the actors (companies and individuals) that are interested in investing money in the Superbonus 110. They deposit fiat money to their client accounts in the Financial Institution and are compensated with the tokens minted in the Investors DAO. 
    \item \textbf{Customers} are the various individuals (companies can not apply to this tax relief program) eligible for the Superbonus. They include natural persons, in relation to expenses incurred for energy efficiency measures carried out on individual property units up to a maximum of two.
    \item \textbf{Financial Institutions (FI)} are the ``qualified'' entities acting as guarantors of the entire system composed by the DAOs. The FIs are banks or registered financial intermediaries, or a company belonging to a banking group that is also registered, as are authorised insurance companies. Their role in our model is to manage both Investors and Operators DAO thus minting, distributing, and burning tokens. They distribute tokens in the Operators DAO and pay back Investors at the end of the program redeeming tokens of the Investors DAO and giving back fiat money.
    
\end{itemize}
 
In the \textbf{Operator group} we have again the Financial Institution and the following actors:

\begin{itemize}
    \item \textbf{General Contractor (GC)} is an operator that acts as a construction company, executing part of the works and as coordinator of other suppliers and sub-contractors. It also acts as a proxy payer and coordinator of Design Architects and Professionals involved in the process. It is in charge of managing all the paperwork involved in the Superbonus 110 like discounted invoicing and tax redeem and transfer processes; consequently, it is the subject that monitors all the payment processes.  
    As construction manager, it shall ascertain and record all events and works generating expenses as soon as they occur so that it can at any moment perform the following activities: (i) Issue the progress statements of the works within the deadline fixed in the tender documentation and in the contract, for the purpose of preparing the paperwork for the advance payments.
    (ii) Monitor the progress of the works and promptly issue the necessary actions for their execution within the limits of the time and sums authorized.
    (iii) Communicate, as construction manager, to the Financial Institution/Investors all the Work Progress States as well as interact with Tax Auditors and Design Architects for certifications of works and fiscal credits matured.

    \item \textbf{Sub-contractor (SC)} is the firm carrying out the works, interacts mainly with the General Contractor both for receiving working orders and for payment issues, all invoicing is done to the General Contractor and doesn't claim credits.
   
    \item \textbf{Supplier} is the entity that supplies materials. It deals with Subcontractors or directly with GCs for the procurement of materials to be used on construction sites.
    
    \item \textbf{Design Architect (DA)} are designers such as engineers and architects. The Superbonus designer must pay special attention to compliance with all sector regulations such as Building Regulations, Landscape Constraints and any other constraints existing in the area. Is in charge of producing all the technical material (e.g., reports and drawings) necessary for the intervention to comply with the regulations in force. They asseverate the technical aspects of the project.
    
    \item \textbf{Tax Auditor (TA).} In the event of professional negligence or misapplication of the rules, the professionals and the technicians in charge of the asseverations and issuing the compliance certificates could be liable for the sums unduly used. In this regard, periodic checks are carried out by a Tax Auditor for compliance with the constraints defined in the smart contracts. He also checks the results of financial transactions, including correct invoicing of expenses (Financial asseveration).
\end{itemize}

\textbf{Roles.} The roles, actions, and related assignments are interdependent elements that help to determine and carry out tasks within an organisation. They define specific tasks and responsibilities, connect them to achieve overall goals, and ensure effective execution. Having identified the actors, we report who/what interacts with the system and with what role in Table \ref{tab:roles-actors}.

\begin{table}[!ht]
\begin{tabular}{|p{0.10\textwidth}|p{0.15\textwidth}|p{0.65\textwidth}|}
\hline
\multicolumn{1}{|c|}{\textbf{Role}} & \multicolumn{1}{c|}{\textbf{Actor}} & \multicolumn{1}{c|}{\textbf{Assignments}} \\ \hline
CCA & TA & Check and control fiscal and financial aspects \\ \hline
CM & GC & Supervises activities on the construction site\\ \hline
DoW & DA & Control expenditures of the works \\ \hline
CW & TA and DA & Check technical and administrative compliance \\ \hline
SG\&S & Supplier, Sub-cont., DA & Supply materials and professional services  \\ \hline

\end{tabular}
\caption{Roles, actions, and related assignments are interdependent elements that contribute to defining and executing tasks in the organisation.}
\label{tab:roles-actors}
\end{table}

\begin{itemize}
    \item \textit{Role 1: Compliance Control Agent (CCA).} [Tax Auditor] This role is connected with the activity of Tax Auditors and comprises all the activities of checking and controlling fiscal and financial aspects. It also certifies all intermediate and final steps of credit maturation and liquidation, it furthermore controls that other professionals' invoices are correct and adherent to price-list approval.

    \item \textit{Role 2: Construction Manager (CM).} [General Contractor] It is responsible for the coordination of all administrative and operational activities to be held on the construction site. This includes also acting as a proxy for payments on behalf of the Customer towards all players operating in the Superbonus process. This a very important coordinating role that must be taken into big account in order to run all the processes in a smooth way.

    \item \textit{Role 3. Director of Work (DoW).} [Design Architect] He has the task of controlling the expenditure connected with the execution of the works, through the accurate and timely compilation of the accounting documents which are public acts to all effects of the law. He also ascertains and registers the facts producing expenditure. The DoW: (i) checks if the work is performed according to the project, (ii) transfers the measurements made to the accounting ledger in order to define the progress of the expenditure.
    He must pay special attention to guarantee compliance with the Superbonus Designer prescriptions, with particular attention to the laying of materials. The Director of Works is thus accountable for the quality and outcome of the work.
    
    \item \textit{Role 4. Checker Work (CW).} [Tax Auditor and Design Architect] During the work and at the end of the construction process, he verifies the quality of the work carried out and issues a report to be sent to the investor. They also asseverate the financial and technical aspects of the works.

    \item \textit{Role 5. Suppliers Goods \& Services (SG\&S).} [Supplier, Sub-contractor, Designer Architect] They are all the subjects that in some way supply materials as well as professional services to Customers and General Contractor. Suppliers are paid by the General Contractor and do not claim tax credits. 

\end{itemize}

\section{Scenarios}
\label{sect:scenarios} 
The process of obtaining the Superbonus 110 reliefs involves working with technical experts, contractors, bankers, and other professionals to plan and carry out the necessary renovation work. In this section, we expose a couple of main scenarios that describe the information flow and the actions related to the aforementioned agents.

\subsection{Scenario1: Homeowner and Technical Expert}
\begin{figure}[!ht]
	\centerline{\includegraphics[width=1.00\hsize]{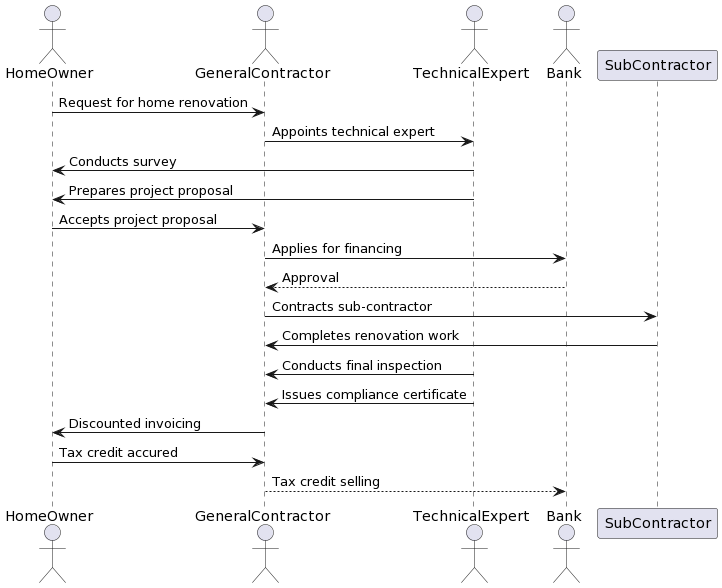}}
	\caption{Sequence diagram of Scenario 1. Interactions between the Homeowner, Technical Expert, Bank, and Contractor in the first scenario are illustrated.} 
	\label{fig:scenario1}
\end{figure}

Agent $a_{ho}$ is a homeowner who wants to renovate his home to improve its energy efficiency and reduce her energy bills as well as the seismic response to earthquakes of the building. She contacts the agent $a_{gc}$, a General Contractor, in order to manage all the aspects of the renovation works. A technical expert $a_{te}$, who is qualified to provide advice on the requirements for the Superbonus 110, is then appointed by the $a_{gc}$. 
Then $a_{te}$ visits $a_{ho}$ home and conducts a survey to identify the work needed to make it more energy-efficient and what anti-seismic improvements can be made. $a_{te}$ then prepares a project that outlines the necessary work and estimates the costs involved. 
$a_{ho}$ accepts the project proposal, then the agent General Contractor $a_{gc}$ applies for financing and credit transfer from his bank and obtains approval. It then contracts an experienced Sub-contractor $a_{sc}$ to carry out the renovation work. After $a_{sc}$ completes the work, $a_{te}$ conducts a final inspection and issues a certificate of compliance. After these passages the $a_{gc}$ is able to invoice and start the discount process in order to redeem the tax credit accrued. All the sequence interactions among the agents are shown in 
figure \ref{fig:scenario1}.


There are four main classes involved in this process: Homeowner, TechnicalExpert, Bank, and Contractor in figure \ref{fig:scenario1classDiagr}. Each class has its own attributes and methods that relate to their role in the process:

\begin{figure}[!ht]
	\centerline{\includegraphics[width=0.60\hsize]{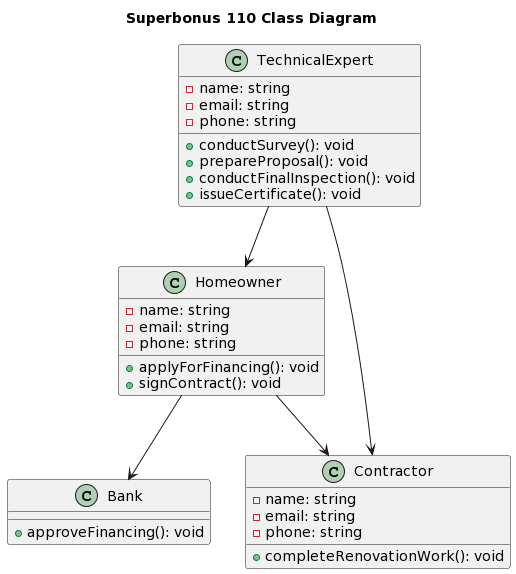}}
	\caption{Class diagram of Scenario 1. Shows the classes involved in the first scenario, including Homeowner, TechnicalExpert, Bank, and Contractor. The classes have attributes such as name, email, and phone, as well as methods such as applyForFinancing(), conductSurvey(), and completeRenovationWork(). The interactions between the classes are shown with arrows and labels.}
	\label{fig:scenario1classDiagr}
\end{figure}

\begin{itemize}
    \item \textit{Homeowner:} The individual who owns the property that will undergo the energy efficiency and/or seismic renovation. The class has attributes such as name, email, and phone, and methods such as applyForFinancing() and signContract().
    \item \textit{TechnicalExpert:} The expert who conducts the energy efficiency survey and provides a project proposal for the renovation work. The TechnicalExpert class has attributes such as name, email, and phone, and methods such as \textit{conductSurvey()}, \textit{prepareProposal()}, \textit{conductFinalInspection()}, and \textit{issueCertificate()}.
    \item \textit{Bank:} The financial institution that provides financing for the energy efficiency renovation. This class has a single method called \textit{approveFinancing()}.
    \item \textit{Contractor:} The company or individual who completes the renovation work. The Contractor class has attributes such as name, email, and phone, and a single method called \textit{completeRenovationWork()}.
\end{itemize}

\subsection{Scenario 2: Financial Institution, General Contractor and Investor}
\label{sec:42}
During the Superbonus 110 reliefs process, the Financial Institution operators manage the Investor DAO ($dao_{IN}$) and Operator DAO ($dao_{OP}$). In this simplified scenario, a single Financial Institution $FI_i$ and a single General Contractor $GC_i$ are considered. $FI_i$ acts as the founder on both DAOs, and it is the guarantor of the tokens $t$ minted in the $dao_{IN}$. 
It is responsible for minting $\mu(t_{dao_{IN}})$ and distribution $\delta(t_{dao_{IN}})$ of minted tokens to investors.) The $dao_{IN}$ is a closed and restricted fund until the end of the program, with a fixed number of investors. Each euro of fiat money invested generates one token $t$ in the $dao_{IN}$. 

The minting phase $\mu(t_{dao_*})$ and burning phase $\beta(t_{dao_*})$ occur in both DAOs, with the freezing phases $\phi(t_{dao_{IN}})$ and minting phases $\mu(t_{dao_{IN}})$ in the $dao_{IN}$. (i) The $\beta(t_{dao_{OP}})$ and release $\rho(t_{dao_{OP}})$ phases occur in the $dao_{op}$, where the tokens are redeemed against the $FI_i$. (ii) In the $dao_{IN}$, the phase of releasing $\rho(t_{dao_{IN}})$ occurs once the credit has matured, and (iii) the $FI_i$ can sell the credit on the $GC_i$. (iv) Finally, the phase of burning occurs in $dao_{IN}$ at the end of the program when the fiat money, including profits, is given back to investors.
An example of such interactions is described in the following paragraph along with corresponding sequence diagrams.

\begin{figure}[ht]	\centerline{\includegraphics[width=1.00\hsize]{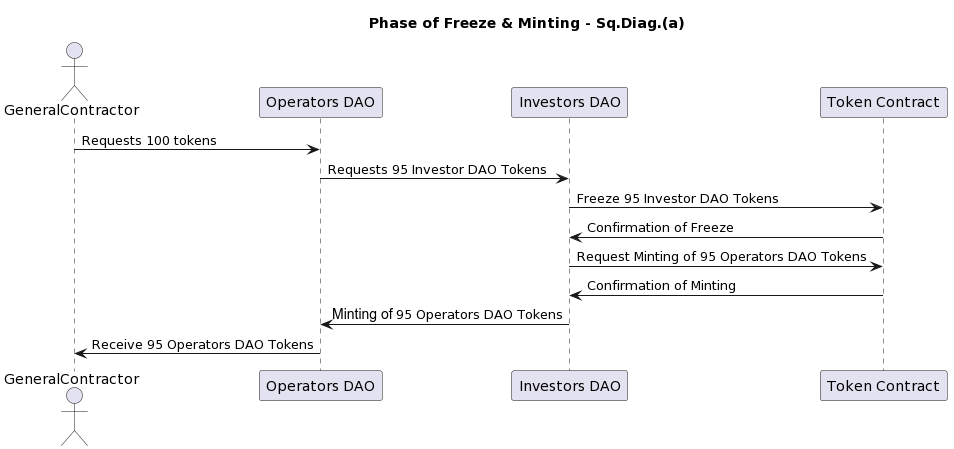}}
\caption{Sequence diagram of Scenario 2. (a) Freeze and Minting in the Investors DAO due to a General Contractor request for a new project.}
\label{fig:scenario2a}
\end{figure}

\begin{figure}[ht]	\centerline{\includegraphics[width=0.95\hsize]{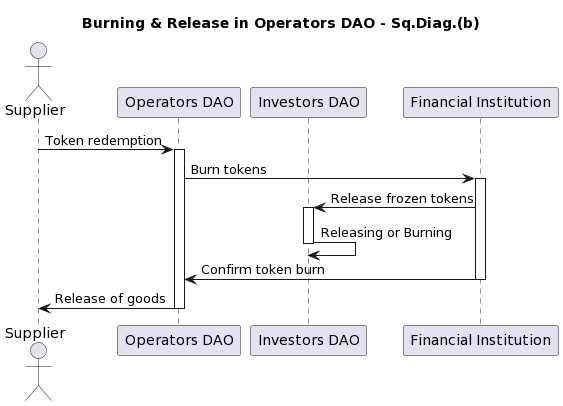}}
\caption{Sequence diagram of Scenario 2 (b) Burn and Release in the Operators DAO following a Fiat liquidation request from Suppliers.}
\label{fig:scenario2b}
\end{figure}

\textbf{a. Phase of Freeze $\phi(t_{dao_{IN}})$ and Minting $\mu(t_{dao_{IN}})$.} Freezing happens in Investors DAO. For every 100 tokens requested from the Operators DAO, 95 Investor DAO Tokens are frozen. Once the link of the 2 tokens is established, Minting takes place: 95 Operators DAO tokens are minted and assigned to the wallet of the requesting General Contractor (see sequence diagram in figure \ref{fig:scenario2a}). 

\textbf{b. Phase of Burning $\beta(t_{dao_{OP}})$ and Release $\rho(t_{dao_{OP}})$.} Once the tokens in the Operators DAO end their life cycle (e.g., arrive at a supplier that needs to monetise the goods supplied) can be redeemed against the Financial Institution. Tokens in the Operators DAO are burned and taken away from the operator wallet, at the same time tokens previously frozen in the Investors DAO are released and are available for Releasing or Burning (see sequence diagram in figure \ref{fig:scenario2b}).

\begin{figure}[ht]	\centerline{\includegraphics[width=0.95\hsize]{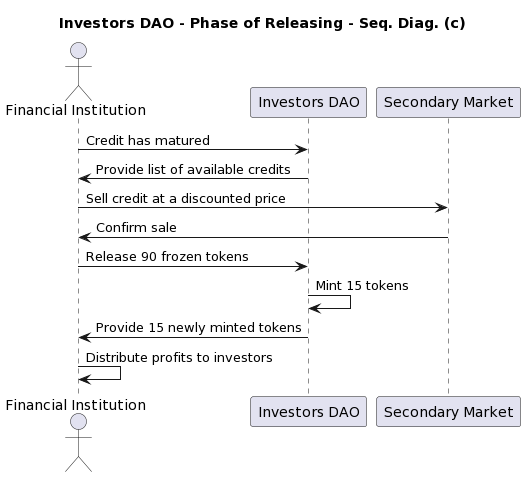}}
\caption{Sequence diagram of Scenario 2 (c) Example of value generation in the Investors DAO due to tax credit selling.}
\label{fig:scenario2c}
\end{figure}

\begin{figure}[ht]	\centerline{\includegraphics[width=0.80\hsize]{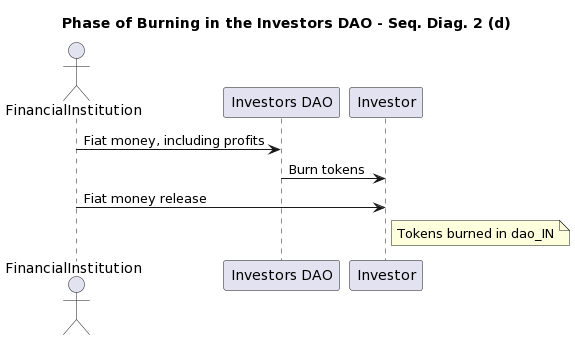}}
	\caption{Sequence diagram of Scenario 2 (d) Profit distribution to Investors and correspondent Investors DAO token burning.}
	\label{fig:scenario2d}
\end{figure}

\textbf{c. Phase of Releasing in the Investors DAO $\rho(t_{dao_{IN}})$.} Once the credit has matured via spending and raise of discounted invoicing (see above phases), the Financial institution can sell the credit on secondary markets. Once this happens, the FI (let’s say 110 euros credit is sold at 105 euros) releases the 90 tokens frozen in the Investors DAO and mints the exceeding 15 tokens generating those profit for the investors (see sequence diagram in figure \ref{fig:scenario2c}).

\textbf{d. Phase of Burning in the Investors DAO $\beta(t_{dao_{IN}})$.}  Tokens are burned in the Investors DAO at the end of the program once fiat money, including profits, are given back to investors (see sequence diagram in figure \ref{fig:scenario2d}).

\section{Secured Fiscal Credit Model (SFCM)} 
\label{sect:sysarc}

The option of a tax credit discount with a benefit bigger than the spent amount is a brand new possibility in the financial and fiscal scenario. It has thus required the definition of a methodology of research capable of evaluating and forecasting new unpredictable situations. Our methodology has pillars that aim to provide a secure and usable environment, such as traceability and immutability of transactions, automated and unbiased control of functionalities, and the ability to forecast risky situations.

In this section, we discuss the option of tax credit discount based on Section \ref{sect:methodology} to evaluate and forecast such unpredictable scenarios. The solution proposed provides tracking functionalities for the entire WPS procedure to protect against risks that the Superbonus beneficiary may incur. An implementation of the Multi-agent system on Algorand Blockchain and using Algorand Smart Contracts for secure fiscal credits DAOs is analysed, where the Investor DAO allows investors to buy credits and benefits from the selling of such credits, while the Operators DAO is a marketplace for financial institutions to distribute fungible tokens to General Contractors.

\subsection{Software Architecture}
\label{sect:architecture}
Our system offers a tracking functionality of the entire WPS procedure to protect against risk situations that the Superbonus beneficiary may incur, such as the liability of professionals, suppliers and assignees in case of damage due to professional negligence, inexperience or misapplication of rules.

The architecture proposed is based on the separation between the Investors' world and the Operators' one, this consents Financial Institutions to raise funds from clients or investors and tokenise such assets. 
The process is designed assuming a simplified model where banks receive transfers of fiat on their accounts and mint the correspondent value of tokens on the Investors DAO. 
 
Another key feature is to guarantee that money anticipation performed by financial institutions and the correspondent underlying credit generation is backed by a fiat asset and is thus repayable when the chain of credit is closed.

\begin{figure}[ht]
	\centerline{\includegraphics[width=1.00\hsize]{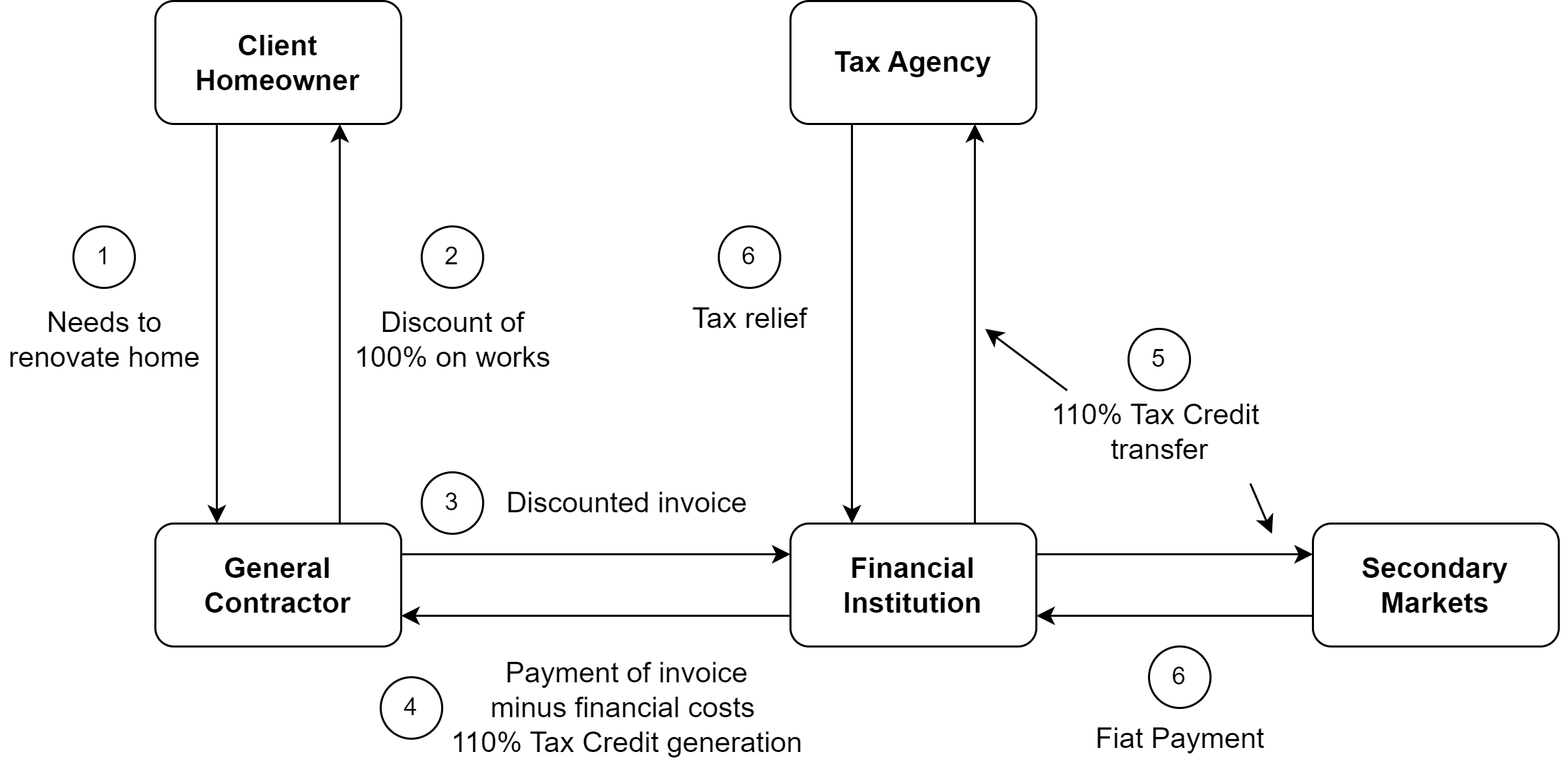}}
	\caption{Cycle of the Invoice Discount Tax Credit generation. At time 5 Financial Institution may decide either to redeem the credit or to use it to pay taxes to Tax Agency.}
	\label{fig:flow}
\end{figure}

\paragraph{Preliminary Assumptions} The option that has been considered and modelled in this paper is the Invoice Discount, as it is the most widely used solution. Unlike direct selling or direct deduction, this option requires credits to be tracked and managed because the benefits may be transferred through many subjects during the generation and redemption process. Another assumption is that the Customers are using a General Contractor to manage the entire process of renovation on the construction site and the associated paperwork. This is to simplify the model and make it easily extendable to a multi-supplier situation. Furthermore, we assume to represent the most common scenario represented in figure \ref{fig:flow}. The General Contractor will be the only party responsible for dealing directly with the customer, coordinating and paying all the sub-contractors, professionals, and suppliers involved in the works.

The system architecture implemented following the preliminary assumptions consists of three main blocks interacting with each other:
\begin{itemize}
  \item A Multi-agent system, consisting of peer-to-peer connected nodes, onto which a decentralised and distributed application based on Blockchain technology is implemented. It will be deployed on Algorand (as outlined in section \ref{sect:algorand}). The system has been designed using the Mesa Framework \footnote{\url{https://mesa.readthedocs.io/en/main/index.html}, last accessed in Feb 2023} an agent-based modelling framework based on Python. This framework provides a modular environment for system visualisation and testing.
  
  \item Two modules consisting of two DAOs based on the Aragon framework (see section \ref{sect:aragon}) used for the development of the system's business logic. Aragon made it possible to manage the communication between the dApp and the Ethereum Virtual Machine in a completely transparent manner.
\end{itemize}

\subsection{Secured Fiscal Credit DAOs}
\label{sec:53}

The Investor DAO allows investors to liaise with Financial Institutions interested in the purchase and resale of credit. The General Contractor sends invoices to the Customer for the works at the end of the SAL (WPS). Therefore it is the General Contractor that interfaces with the Investor DAO becoming the collector of all credits accrued within the scope of the works. The Financial Institute regulates all monetising because it receives fiat currencies from investors and must guarantee the transparency of financial transactions and the licence to collect funds. At the end of the investment cycle, there will be $n$ investors who have issued fiat currencies and received in proportion to the investment $n$ tokens.

In brief, this DAO is meant to represent a simplified investing fund where participants get tokens proportionally to their investment, keep them blocked until the end of the fund duration, and get a reward proportional to their initial quota at the end. Here we have Fungible Tokens that are minted upon fiat deposit or credit selling by the Financial  Institution and freeze or unfreeze when a guarantee from Operators DAO is requested or released.

\begin{figure}[ht]
	\centerline{\includegraphics[width=1.00\hsize]{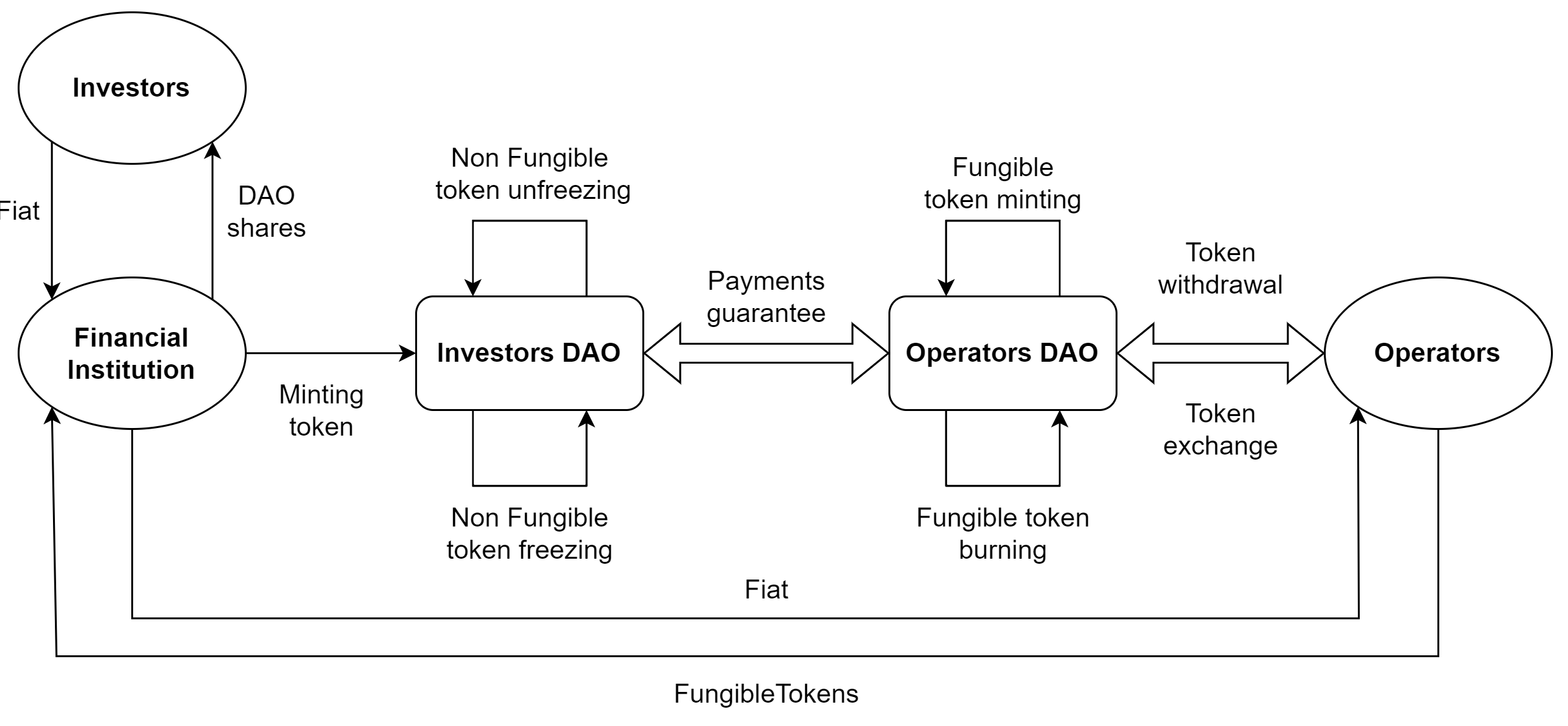}}
	\caption{DAOs Interaction Diagram. The architecture can be seen as a Distributed Trust and Reputation Management System (DTRMS) implemented through two DAOs. Fungible and Non-Fungible Tokens life-cycle is also described.}
\label{fig:arch}
\end{figure}

The \textbf{Investors Group} includes all actors getting financial burdens and benefits in the process of Superbonus 110. The \textbf{Investors} belongs to the Investors Group, which invests money to buy credits and benefits from their selling, while all the other operators General Contractors, Sub-contractors, Suppliers, Design Architects, and Tax Auditors belong to the Operators Group, which are hired and paid by the General Contractor. 

The gain of the investors is proportional to their quote in the DAO. Let's assume $X$ as a set of investors $x_i \in X$, with $|X| = n$. The investor's share $sh_{x_i}$ is equal to the sum of tokens $\sum t_{dao_{IN}}$ of $x_i$ at the beginning of the process. Given $T'$ the sum of tokens in the Investor DAO at the starting time $t_0$, and $T''$ the sum of tokens in the Investor DAO at the end time $t_n$ of the Superbonus procedure, we define with the earn $E_{x_i}$ for an investor, equal to:

$$E_{x_i} = (T'' - T') \times \frac{sh_{x_i}}{\sum_{i=1}^n sh_{x_i}},$$

where $T'' - T'$ is the total earning in term of tokens, and $\frac{sh_{x_i}}{\sum_{i=1}^n sh_{x_i}}$ is the percentage of share held by $x_i$ at time $t_0$.

The second DAO is a marketplace for Financial Institution to distribute fungible tokens to General Contractors, these are minted and burned in order to satisfy credit requests from Operators and spent for buying goods and services. Each minting on Operators DAO corresponds to a token freezing of the same amount on the first DAO in order to guarantee the coverage. The tokens are redeemable by the Financial Institution which is also the guarantor of the whole system. Once the second DAO tokens are redeemed, the first DAO NFTs are unfrozen. Each fungible token minted here is coupled via a unique code to the tax credit generated by that spending of money.
A scheme of interaction among DAOs and Actors is shown in figure \ref{fig:arch}.

\paragraph{Tokenomics in the DAOs environment}

The \textbf{Community} that makes up our system consists of the Actors listed in Section \ref{sect:32}. The interaction among them initiates the \textbf{Token Distribution} that takes place in DAOs operating as a closed fund. In particular, the Investors DAO will, as a first step, collect all the fiat funds coming from Investors and convert them into local tokens. At this moment must be stated all relevant financial information and the duration of the operation (which will be connected to the end of government incentives).
The assigned \textbf{Governance Tokens} are used both as voting power rights in the DAO and minting tokens means creating tokens, either from scratch at the beginning of the system or increasing the supply to add more later in relation to credit transactions. 
On the Investor DAO, a token is created for every euro paid in. Freezing is performed each time minting takes place on the Operator DAO (i.e. an operator's request for monetising of accrued credit). Burning happens at the end of the DAO when investments are redeemed.
Token minting means creating tokens, either from scratch at the beginning of the system or increasing the supply to add more later on as a function of credit transactions. On the Investor DAO, a token is created for every euro paid in. Burning is performed each time minting takes place on the Operator DAO (i.e. an operator's request for monetising of accrued credit).
The \textbf{Economic Dynamics} lives around the tokens minted in the Investor DAO, which are issued exclusively to the GC. The GC can execute transactions on its own as the customer's contact person. 

On the \textbf {Investors DAO}, at the closing of the fund, the FI will convert the tokens accumulated in Investors' wallets into fiat and redistribute the shares to them, including dividends. The excess tokens will be generated, on one hand by the disposals towards the Operator DAO, which considered 100 the amount of the work will receive a lower share of tokens for example 95 to compensate for the financial costs. On the other hand, the 100 tokens will generate 110 tax credits that will be sold by the FI on secondary markets to obtain for example 105. Thus the investors for an investment of 100 will receive 115 (see section \ref{sect:scenarios} on token burning and minting).

On the \textbf {Operators DAO}, for example, the GC receives 100 tokens of which 30 go to the Supplier, 20 to the Design Architect, and 10 to the Tax Auditor for audits based on invoices they have issued. He can make a profit of 40 tokens for the operation. All these tokens are redeemable into fiat either upfront or after a determined time to consent of the Financial Institution to organise payments.

\subsection{A framework for DAOs} 
\label{sect:aragon}
The framework chosen to develop the DAOs modules of the demonstrator is Aragon\footnote{\url{https://aragon.org}, last access Feb 2023}. This is a second-level platform based on the Ethereum network that is natively designed to model government systems of public and private entities, it is easy to use and allows to deploy of test applications at very cheap transaction costs using Ethereum's test nets such as Rinkeby\footnote{\url{https://www.rinkeby.io}, last access Feb 2023}. 

It is furthermore an independent platform oriented to design and implement DAOs and is thus particularly suitable for creating configurable governance structures. Aragon is ultimately a software for creating and governing organisations such as companies or cooperatives\footnote{\url{https://documentation.aragon.org/aragon/readme}, last access Feb 2023}.
The platform is based on the Aragon Network Token (ANT) which grants voting rights to its holders, polls are used to make decisions on the development of a DAO within Aragon.
The model described is implemented through several smart contracts written with Solidity\footnote{\url{https://soliditylang.org}, last access Feb 2023} an object-oriented language specifically meant for contract implementation. The choice of such stiff language is because we want the execution to be strictly what is supposed to be without the possibility of modifications. 

\textbf{A sample application on Aragon.} We have implemented a demonstrator on Aragon to simulate the operating principles of Investors and Operators DAO, in a future development the integration with this demonstrator and the one described in this article could be implemented. The Aragon demonstrator along with a representation of Smart contract's states and definition of global and local constrain, data structures and actors involved is illustrated in \cite{DeGasperis2022Facchini}.

\subsection{SFC Multi-agent System}
\label{sect:SFCMAS}
The MAS module of the demonstrator is composed of two main sub-systems, one internal meant to manage the core functions of the Superbonus110 processes, the second external to interface the whole MAS to the Double-DAOs module.

\textbf{MAS-DAO Communication.} The interaction with DAOs happens through two service agents whose functionality is to interface Investors and Operators DAO with the MAS, they supervise all information on basic operations coming from/to the DAOs. This part is not implemented at the moment in the demonstrator.

\textbf{Internal MAS.} Each agent has the relevant logic and internal communication functions integrated, in this way, blockchain interactions happen on a singular base and are limited to what the agent's needs are. This part is implemented and described in section \ref{sect:algorand}.

Basic features of the MAS module are realised through several agent categories to model all the roles defined in the previous section at Table \ref{tab:roles-actors}. 
General Contractors, Subcontractors, Suppliers, Design Architects, Tax Auditors are categories that are pretty similar operating basically as negotiating agents. They are defined by homogeneous data (registry data) and similar functions to perform (checking payments received, verifying docs exchanged, etc...). This agent will react to human requests (e.g., approve a credit certification) and to other agents' requests. Clients (homeowners) have the initial function to hire General Contractors and Technicians while  Financial institutions fund General Contractors once they are appointed by the Client. A particular role instead is given, in our system, to the Workflow Agent a special agent used to model paperwork and evaluate advances of the works. There are five states modelled (Open, Anticipation, Sal1, Sal2, Eow) plus the final one for archiving (Archived) as per workflow described in section \ref{sb110}.
The relevant features of agent classes will be as follows:

\begin{itemize}
  \item \textbf{Workflow Agent} is used to represent each Superbonus 110 process opened. Once the instance is created the agent will be on position $(x,0)$, as soon as the work is approved will move to position $(x,1)$ corresponding to SoW state and it will move as soon as the state passage condition is met (e.g., related Tax Auditor agent approve the credit for that stage) until it reaches the final position $(x,5)$. Once the paperwork is done the agent is archived (possibly for 7 years as per law requirements). 
  The agent logic thus evaluates if the represented paperwork is due to be moved to the next state, i.e. the Technicians have asseverated the current SoW and the General Contractor has paid the anticipation for the next state. When the agents have instantiated the total value of works is generated with a random procedure with a value comprised between 1 mln and 10 mln microAlgos. 
  
  We modelled, to avoid excessive complexity, 5 paperworks to be handled in the demonstrator, thus having 5 Workflow agents.
  The logic is realised through a smart contract deployed on the Algorand testnet wallets of agents and is activated by opting into such a smart contract. Considering the current state of the Workflow agent as $x_{cur}$, the last asseverated state as $x_{asv}$ and the last paid as $x_{pay}$, the propositional logic notation for such constraint evaluation is the following:
  
  \begin{center}
  $ x_{cur}(i+1) \Longleftarrow \neg (x(i)_{cur} = x_{asv}(i)) \land (x_{asv}(i)=x_{pay}(i)) \land \neg x_{arc}(i).$
  \label{eq:paperwork} 
  \end{center}
  
  The Workflow Agents also pay the Technician agents once the state is successfully updated, this happens by sending between Algorand wallets the amount calculated as a percentage (adjustable from the specific home page slider) of the state value (adjustable again from relevant sliders), see figure \ref{fig:frontend} on Section \ref{sect:MASmodelling}.

  \item \textbf{General Contractor Agent} has the function of approving, on the financial side, the evolution to the next step of the paperwork. It simulates the passage of state generating a random number to be checked against a threshold configurable as an input parameter in a similar way to the Workflow agent's procedure. If the approval is successful, the agent sends from his wallet on Algorand the amount due as anticipation for the next state to the related Workflow agent. The amount is configurable setting the percentages of each state against the total value of the paperwork. To keep the system simple we modelled a single General Contractor Agent.

  \item \textbf{Technical Agent}: This class merges the functionalities of both Design Architects and Tax Auditors with approvals randomly generated and checked against thresholds adjustable. In our simulator, we modelled two technician agents, where each agent has an Algorand wallet to get payments for asseverations.
  
  \item \textbf{Financial Agent} represents the Financial Institution and acts as a sort of notary and banking entity where all other General Contractors agents refer for funding approval, payment requests, and other Superbonus 110-related financial issues. This agent also interacts with Communication Agents to perform all the token operations between the DAOs, but will be modelled in the next version of the demonstrator.

  \item \textbf{Client Agents}: Clients, model the homeowners activating Superbonus 110 interventions, appoint the General Contractor and Technicians executing all the relevant works and certifications required by law. Following the condominium assembly, it creates the Workflow and assigns the aforesaid agents to it. It's not modelled in this version of the demonstrator and the General Contractor and Technicians assignment to Workflow agents is hard-coded.
 
\end{itemize}

\begin{figure}[!ht]
	\centerline{\includegraphics[width=1.00\hsize]{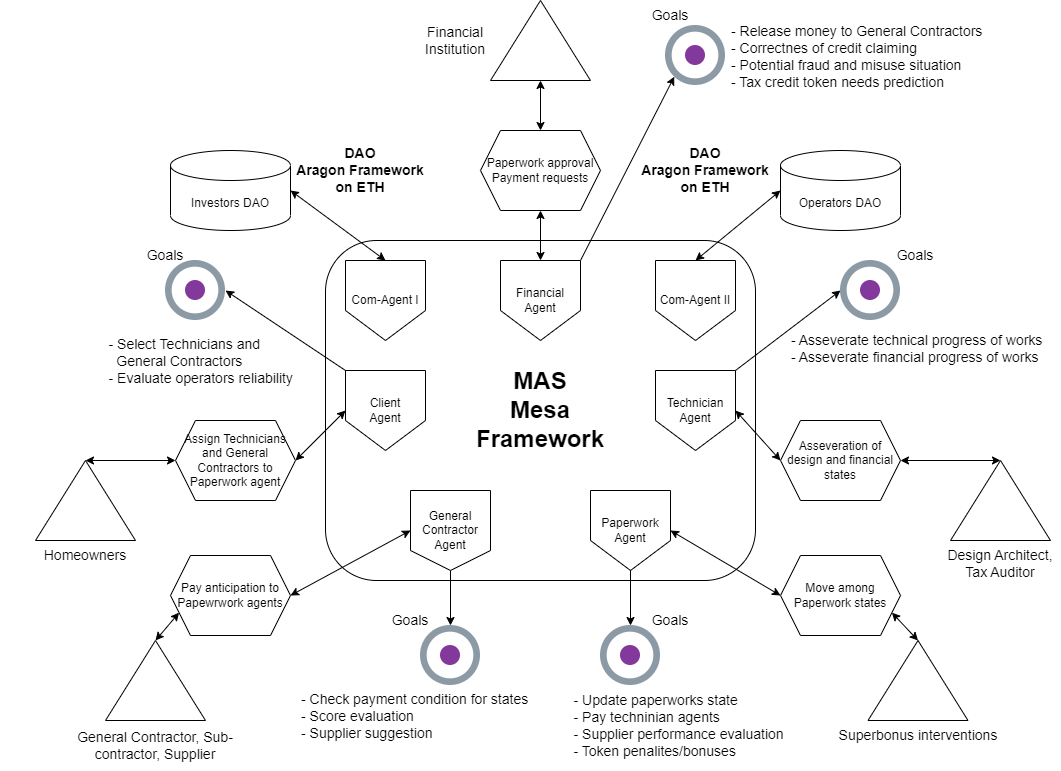}}
	\caption{Multi Agent System architecture. The demonstrator we have implemented is meant to illustrate the benefits of introducing the DAOs and MAS to Superbonus 110 environment.}
\label{fig:mas}
\end{figure}

A common feature to agents is a wallet on the selected blockchain to perform and receive relevant payments and token operations or other service activities.
For the SFC architecture design, we used the MESSAGE/UML paradigm to model the agent interaction and behaviour \cite{Ozkaya} in figure \ref{fig:mas}.

\paragraph{Advanced features of MAS Module} Coming to the \textbf{Intelligent aspects} of the system, in addition to previous features, the agent will also have the following properties:

\paragraph{Property 1}[Workflow Agent] In order to increase the performance of the operative players of the  Superbonus 110 environment, the Workflow agent evaluates the performance of its suppliers. The penalties and rewards are based on parameters like response time to requests, discounts on tariffs, and respect for delivery deadlines. Thus at each WPS step the agent evaluates the bonus or malus of each Supplier involved in its construction site modifying his score. It also applies a token penalty to "bad" agents that can be used to reward "good" agents. In this way Operators not performing well have to increase performance in order to avoid losing reputation and money.
Consider $t_{avg}(GC_{x_{i}})$ the historical average time for General Contractor GC to complete WPS states $x_{i}$, $d(GC_{x_{i}})$ the actual discount proposed for such state and $p_{avg}(GC_{x_{i}})$ the historical percentage of on-time WPS completed. We can calculate a weighted value $$w(GC_{j})=w_1*t_{avg}(GC_{x_{i}})+w_2*d(GC_{x_{i}})+w_3*p_{avg}(GC_{x_{i}})$$ of such parameters for supplier $GC_{j}$ and check it against a threshold limit value l in order to determine if a supplier is "good" or "bad":
$$ w(GC_{j})\leq l $$

\paragraph{Property 2} [Financial Agent] This functionality is a credit anomalies detector that checks anomalous credit transfers. The idea is to spot and evaluate possible fraud and misuse situations and prevent them from damaging Customers, Financial Institutions and Operators. It enriches the Constraints 2, 3 and 4 of section \ref{sub:CKB} and is realised through a more complex analysis of token movements between parties such as too-fast token redeem, meaning works claimed could be fake. If general contractor $GC$ ask for a SAL credit redeem having performed WPS states $x_{1}$ and $x_{2}$ in time  $t_{GC} = t(GC_{x_{1}}) + t(GC_{x_{2}})$ then we can spot a suspicious behaviour of $GC$ if 
$$ t_{GC}\leq s * [t_{avg}(GC_{x_{1}}) + t_{avg}(GC_{x_{2}})]$$
 where s is a suspicion rate percentage to be tuned with data from simulation (e.g. if s=0.5 means claims done in half the normal time are considered suspicious).
 
\subsection{MAS Framework and Implementation}
\label{sect:MASmodelling}
\begin{figure}[!ht]
\centerline{\includegraphics[width=1.00\hsize]{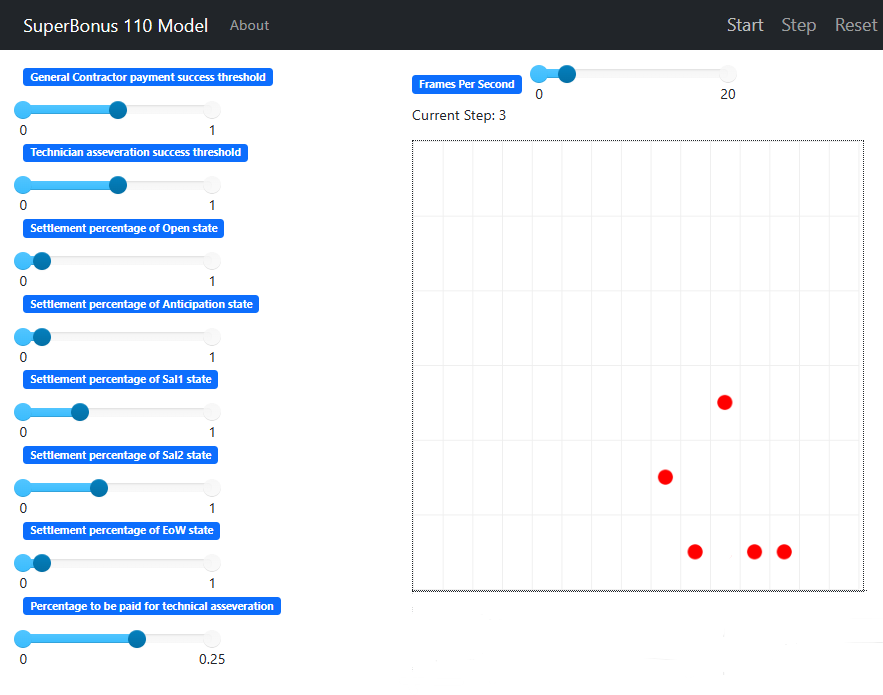}}
\caption{Multi Agent System front end. The grid shows Workflow agents moving vertically as their states get asseverated. On the left, a group of sliders allows to set simulation parameters, while on the header simulation Start/Stop commands are present.}
\label{fig:frontend}
\end{figure}

The solution used to implement the Multi-agent system is the framework MESA, a modular framework in Python language. MESA allows to simulate the complex social systems by creating agents for each actor and designing their interactions in an environment. MESA framework includes various tools and modules for building, running, and analysing MAS models, such as built-in agent behaviours, scheduling mechanisms, data collector primitives, and visualisation tools (see \cite{masad2015mesa}).
Each of the agents' categories described in section \ref{sect:SFCMAS} is modelled in the system. 
We defined different types of agents, such as reactive and deliberative agents, and the mechanisms for implementing communication and coordination among them. The graphical representation of the agents is defined in a separate file named server.py that contains all the visualisation instructions, the input data to be collected (in our case through configurable sliders), and the format of the output. Our demonstrator consists of a web page on port 8521 executed in localhost as in figure \ref{fig:frontend}, generated customising Mesa primitives and showing agents moving on a grid (each agent moves one step vertically once a new state is completed). The simulation can be started and stopped by the relative commands on the top right header, Step allows to run on step per click (a tick is given to the Mesa framework).

\subsection{Blockchain Framework}
\label{sect:algorand}
 As said before Mesa MAS environment is integrated with Algorand, this solution allows transferring the benefits of blockchain in terms of trust and traceability to the whole system.
 For the purpose of the research, we have deployed on the Algorand testnet an account for each agent. Each account is both a wallet to collect and pay fees and an application area to upload smart contracts.
 The environment is set up on a Windows 11 machine through a WSL GNU/Linux virtual machine and Docker Desktop. The tool used for setting up the testnet is Algorand Sandbox and its three docker containers (indexer, Postgres and algod). The accounts are manually setup and consent to both to open a wallet, used to simulate payments through the various agents and to deploy applications.
To implement, test and deploy the smart contract on the Workflow agents we instead used Algokit Sandbox\footnote{\url{https://developer.algorand.org/docs/get-started/algokit}, last access Feb 2023} with Dappflow\footnote{\url{https://dappflow.org}, last access Feb 2023}.

\begin{figure}[!ht]
	\centerline{\includegraphics[width=1.00\hsize]{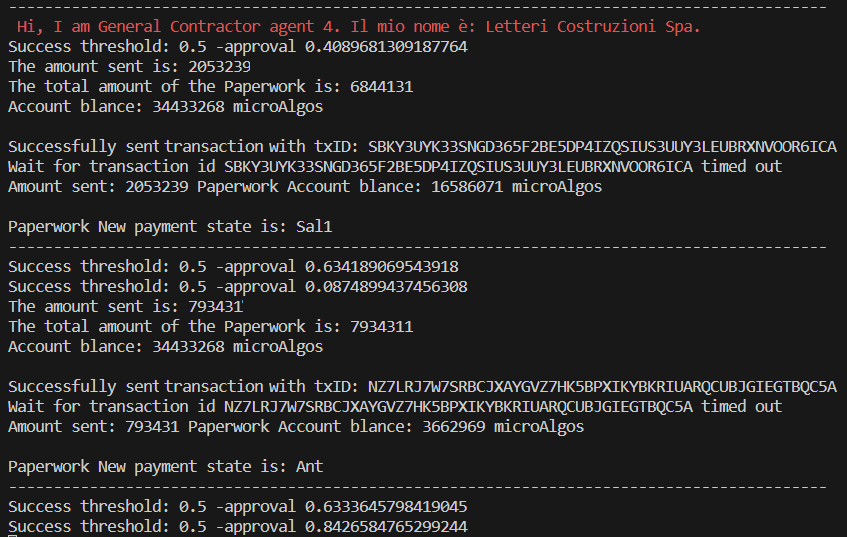}}
	\caption{General Contractor agent behaviour. The agent checks if the randomly generated approval value is smaller than the threshold, and if true pays to the Workflow agent's wallet the amount calculated as the current state percentage of the total value.}
\label{fig:generalcontractor}
\end{figure}

An example of General Contractors output is illustrated in figure \ref{fig:generalcontractor}
Each agent has an ALGO-wallet connected to the Vault located in the testnet \footnote{\url{https://testnet.algoexplorer.io}, last access Feb 2023}.
    
\textbf{Algorand Smart Contracts} are pieces of logic remotely callable that reside on the Algorand blockchain, these contracts are primarily responsible for implementing the logic associated with a distributed application. Smart contracts can generate asset and payment transactions on the blockchain defined using an assembler-like language called \textit{Teal}. In our demonstrator, the smart contracts are written in Python and integrated into Algorand using PyTeal\footnote{\url{https://github.com/algorand/pyteal}} libraries and deployed to the blockchain using the Algokit SDK. Another tool used is Beaker \footnote{\url{https://github.com/algorand-devrel/beaker}} a development environment to test and deploy Pyteal programs.

\section{Conclusions}
\label{sect:concl}
The present work is an extension of our previous publication \cite{DeGasperis2022Facchini} presented to PAAMS 2022 Conference and of the abstract presented to Doctoral Consortium of IJCAI 2022 \cite{ijcai2022p828}. We have improved the modules of the Multi-agents area defining the Knowledge Base constraints and implementing a web-based software to simulate the interaction of a set of Superbonus 110 players.
We also detailed with more importance both the DAOs properties and the Intelligent part of the system implementing the number and the functions of agents and suggesting a potential area of investigation to develop an intelligent MAS controlling. In particular in Section \ref{sect:SFCMAS} we defined two properties to spot fraudulent behaviour of agents and a reward system to encourage good Suppliers.

The model used in our demonstrator could be easily exported to other areas both in the institutional and private sectors. In particular, Public Administrations may benefit from a reliable governance and voting system to provide services that require tracking of items and documents (e.g., Board resolutions, Certifications, etc.) \cite{Chohan}. In general, every process where a tracking of some as-set or document is required could benefit from the architecture of Blockchain in the environment of Distributed Trust and Reputation Management Systems(DTRMS) \cite{Bellini}.

In the future, our work will focus on two aspects to improve SCFM: (i) train Machine Learning models to conduct automated smart contract intent detection \cite{Huang2022Youwei}; (ii) use a client-side validated state like RGB\footnote{\url{https://www.rgbfaq.com} Last accessed August 2023} a suite of protocols for scalable and confidential smart contracts system operating on Layer 2 and 3 of Bitcoin ecosystem, and lightning network. Another possible future development may consider the integration of smart contracts and agents with IoT solutions such as sensors to monitor physical quantities.

\bibliographystyle{unsrt}  
\bibliography{references}

\end{document}